\documentclass[a4paper,12pt]{article}
\usepackage{jheppub}

\pdfoutput=1
 \usepackage{hyperref}
\hypersetup{colorlinks=true,linkcolor=magenta,anchorcolor=green,citecolor=cyan,filecolor=black,menucolor=black,urlcolor=black}
\usepackage{graphicx} 
\usepackage[table]{xcolor}
\usepackage{graphicx}
\usepackage{amsmath,amssymb}
\usepackage{soul}

\newcommand{\be}{\begin{equation}}
\newcommand{\ee}{\end{equation}}
\newcommand{\bea}{\begin{eqnarray}}
\newcommand{\eea}{\end{eqnarray}}

\newcommand{\bdm}{\begin{displaymath}}
\newcommand{\edm}{\end{displaymath}}

\definecolor{bubbles}{rgb}{0.91, 1.0, 1.0}
\definecolor{aquamarine}{rgb}{0.5, 1.0, 0.83}
\definecolor{bubblegum}{rgb}{0.99, 0.76, 0.8}
\definecolor{bluebell}{rgb}{0.64, 0.64, 0.82}
\definecolor{dollarbill}{rgb}{0.72, 0.93, 0.6}

\usepackage{tikz,fp}
\usetikzlibrary{external,fixedpointarithmetic,decorations.pathmorphing,positioning,calc,fadings,decorations.markings,decorations.pathreplacing,patterns,arrows}

\tikzset{
  label distance=-3pt,
  >=stealth,
  inner sep=1.5pt,
  boson/.style={
    decoration={snake, segment length=2mm, amplitude=0.5mm},
    decorate,
  },
  scalar/.style={
    dashed
  }
}

\begin{document}
\preprint{IPHT-t24/012}
\title{Schwarzschild geodesics from Scattering Amplitudes to all orders
  in $G_N$}
\author{Stavros~Mougiakakos$^{(1)}$ and Pierre Vanhove$^{(2)}$}
\affiliation{(1) Laboratoire Univers et Th\'eories, Observatoire de Paris, Universit\'e PSL, Universit\'e Paris Cit\'e, CNRS, F-92190 Meudon, France}
\affiliation{(2) Institut de Physique Th\'eorique, Universit\'e  Paris-Saclay, CEA, CNRS, F-91191 Gif-sur-Yvette Cedex, France}

\date{\today}

\abstract{The dynamics of the leading self-force (0SF) order, corresponding to the geodesic motion of a light body in the exact background of a heavy body, are explicitly derived for the first time using a flat space scattering amplitude-based approach. This is achieved by utilising the cubic formulation of Einstein's general relativity coupled to the effective worldline action of massive point particles, which was employed to derive the Schwarzschild black hole metric in~\cite{Mougiakakos:2024nku}.
}

\maketitle

\section{Introduction}

The dynamics of gravitational spinless binary systems depend on three classes of physical parameters: the relative velocity between the bodies in question, the mass ratio of the two bodies in question and the strength of the gravitational force.

In the weak field regime at large impact parameters, a perturbative approach using the formalism of scattering amplitudes on a flat spacetime in a post-Minkowskian expansion can be employed~\cite{Bjerrum-Bohr:2013bxa,Cheung:2018wkq,Bern:2019nnu,Mogull:2020sak,Kalin:2020mvi,Kalin:2020fhe,Dlapa:2021npj,Bjerrum-Bohr:2021din,Bern:2021dqo,Bern:2021yeh,Herrmann:2021tct,DiVecchia:2021bdo,Dlapa:2022lmu,Kalin:2022hph,Bjerrum-Bohr:2022ows,Damgaard:2023ttc,Georgoudis:2023eke,Georgoudis:2024pdz,Bini:2024rsy,Driesse:2024xad}.  
This weak-field approach is for generic masses of the binary system and all regime of relative velocities. 
The strong-field regime can be studied using  the self-force  formalism (see~\cite{Poisson:2004gg,Wald:2009ue,Barack:2018yvs} for reviews from a general relativity perspective). This is an expansion in the mass ratio of the binary system but to all order in gravitational strength $G_N$. 
Damour observed~\cite{Damour:2019lcq} that the polynomial dependence of post-Minkowskian results on the masses allows for a comparison with the self-force expansion. By expanding the self-force results, one can connect with the post-Newtonian expansion~\cite{Blanchet:2010zd,Barausse:2011dq,Shah:2013uya,Damour:2015isa,Bini:2017wfr,Long:2024ltn}
or post-Minkowskian results~\cite{Bini:2019nra,Damour:2019lcq}  for bound orbits.
A recent development has been the initiation of an effort to compute
gravitational dynamics in the self-force expansion by combining the perturbative scattering amplitudes approach with curved-space methods~\cite{Driesse:2024xad,Cheung:2020gyp,Cheung:2020gbf,Kosmopoulos:2023bwc,Cheung:2023lnj,Cheung:2024jpo,Adamo:2023cfp}.  Most of the existing approaches assume  the geodesic motion of the light body, and build on top of that towards  deviations to the geodesic motion. The scope of this letter is to present, for the first time, the exact derivation of the geodesic motion using a flat spacetime scattering amplitude approach. This is a first step in setting a self-force analysis to all orders in $G_N$ from an   amplitude perspective.

\medskip 

In this work we explicitly derive the geodesic equation for a light body of mass $m$ in the exact curved  Schwarzschild spacetime created by a heavy body of mass $M$, in the context of weak field perturbation by summing to all orders in $G_N$.
We work with  the cubic formulation of
Einstein's gravity~\cite{Cheung:2017kzx} that was used in~\cite{Mougiakakos:2024nku} to derive the exact
Schwarzschild metric generated by massive spinless pointlike
object.  To this end, we introduce an effective
field theory for the self-force dynamics of the light body in
section~\ref{sec:sfeft}. Then in section~\ref{sec:geodesic}, we  derive the geodesic motion from
summing two infinite perturbative contributions:  one infinite sum
from the black-hole metric generated by the heavy body and one
infinite sum for the couplings between the light body and the
curved background. Section~\ref{sec:conclusion} contains our conclusions.

\section{A self-force EFT formalism}\label{sec:sfeft}

In this section, we will briefly present a self-force effective
field theory for General Relativity.
We  consider a binary system of a heavy body of mass $M$ and a light
body of mass $m$ interacting gravitationally. The 
dynamics is described by~\cite{Kalin:2020mvi,Mogull:2020sak,Cheung:2023lnj,Kosmopoulos:2023bwc,Cheung:2024jpo,Goldberger:2004jt}
\begin{equation}
\mathcal S=\mathcal S_{EH}+\mathcal S_l+\mathcal S_H,    
\end{equation}
composed of  the Einstein-Hilbert action
\begin{equation}\label{e:EH}
\mathcal S_{EH}=\frac{1}{16  \pi G_N}\int d^Dx\ \sqrt{-g} \mathcal R(g), 
\end{equation}
the worldline action for the light  body
\begin{equation}\label{e:Sldef}
\mathcal S_l=-\frac{m}{2}\int d\tau_l\
                      \left(e_l^{-1} \,
                        g_{\mu\nu}v_l^{\mu}v_l^{\nu}+e_l\right)    
\end{equation}
and the worldline action for the heavy body
\begin{equation}\label{e:SHdef}
\mathcal S_H=-\frac{M}{2}\int d\tau_H\
                      \left(e_H^{-1} \,
                        g_{\mu\nu}v_H^{\mu}v_H^{\nu}+e_H\right)   . 
\end{equation}

We use the same cubic formulation of gravity  described
in~\cite{Mougiakakos:2024nku,Cheung:2017kzx} which we briefly recall.
With the introduction of an auxiliary field $A^a_{bc}$ the
Einstein-Hilbert action takes the form
\begin{equation}
 16  \pi G_N \mathcal{S}_{EH}=-\int d^Dx \left(\left(A^a_{bc}A^b_{ad}-\frac{1}{D-1}A^a_{ac}A^b_{bd}\right)\mathfrak{g}^{cd}+A^a_{bc}\partial_a\mathfrak{g}^{bc}\right),
\end{equation}
where $A^a_{bc}=A^a_{cb}$ is an auxiliary field
and $\mathfrak{g}^{ab}=\sqrt{-g}g^{ab}$ is the \textit{gothic inverse
  metric}. We add the gauge fixing term 
\begin{equation}
\mathcal{S}_{GF}=-{1\over 32\pi G_N}\ \int d^Dx\ \eta_{cd}\partial_a\mathfrak{g}^{ac}\partial_b\mathfrak{g}^{bd},
\end{equation}
which correspond to the harmonic gauge condition
$\partial_a\mathfrak{g}^{ab}=0$ for the \textit{gothic inverse
  metric}. We work perturbatively by expanding the \emph{gothic inverse metric}
near flat space 
\begin{equation}\label{e:pertgoth}\mathfrak{g}^{ab}=\sqrt{-g}g^{ab}=\eta^{ab}-
  \sqrt{32\pi G_N}  h^{ab}.
\end{equation}
We will use the mostly positive metric signature $(-,+, \dots, +)$.
The interactions between the auxiliary field and
the metric perturbations are decoupled after shifting 
the auxiliary field as
\begin{equation}\label{shift}
A^a_{bc}\rightarrow A^a_{bc}-\frac{\eta^{ad}}{2}\left(\partial_{(b}h_{c)d}+\frac{\eta_{bc}\partial_d h}{D-2}-\partial_d h_{bc}\right).
\end{equation}

In this formalism, using the metric degrees-of-freedom
$g^{\mu\nu}=\mathfrak{g}^{\mu\nu}/(\sqrt{-\mathfrak{g}})^{2/(D-2)}$
and setting the einbein to 1,   the  worldline actions for the
light body of mass $m$ in~\eqref{e:Sldef} becomes
\begin{equation}\label{e:Slight}
\mathcal S_l= -\frac{m}{2}\int d\tau_l\ \left(\frac{\mathfrak{g}^{\mu\nu}v_{\mu}v_{\nu}}{\left(\sqrt{-\mathfrak{g}}\right)^{\frac{2}{D-2}}}+1\right),
                    \end{equation}
and for the heavy body of mass $M$ in~\eqref{e:SHdef} becomes
\begin{equation}\label{e:Sheavy}
\mathcal S_H= -\frac{M}{2}\int d\tau_H\ \left(\frac{\mathfrak{g}^{\mu\nu}v_{H\mu}v_{H\nu}}{\left(\sqrt{-\mathfrak{g}}\right)^{\frac{2}{D-2}}}+1\right).
\end{equation}

From the above, we can derive the worldline $n$-graviton Feynman rules as
\begin{align}
t^{\alpha_1\beta_1,\dots,\alpha_n\beta_n}_{l\,(n)}&=\left(32\pi
  G_N\right)^{n/2}\frac{i m}{2}v_{l\mu}v_{l\nu}\mathcal{T}^{\mu\nu\alpha_1\beta_1,\dots,\alpha_n\beta_n}_{(n)},\cr
t^{\alpha_1\beta_1,\dots,\alpha_n\beta_n}_{H\,(n)}&=\left(32\pi
  G_N\right)^{n/2}\frac{i M}{2}v_{H\mu}v_{H\nu}\mathcal{T}^{\mu\nu\alpha_1\beta_1,\dots,\alpha_n\beta_n}_{(n)},
\end{align}
where we have introduced the tensor
\begin{equation}\label{e:proj}
\mathcal{T}^{\mu\nu\alpha_1\beta_1,\dots,\alpha_n\beta_n}_{(n)}=\eta^{\mu\alpha_n}\eta^{\nu\beta_n}\mathcal{P}_{(n-1)}^{\alpha_1\beta_1,\dots,\alpha_{n-1}\beta_{n-1}}-\eta^{\mu\nu}\mathcal{P}_{(n)}^{\alpha_1\beta_1,\dots,\alpha_n\beta_n}, 
\end{equation}
where we used the expansion
\begin{equation}\label{e:proj2}
\frac{1}{\left(\sqrt{-\mathfrak{g}}\right)^{\frac{2}{D-2}}}=1+\sum_{n=1}^{\infty}(32\pi
G_N)^{n\over2}\,\mathcal{P}_{(n)}^{\alpha_1\beta_1,\dots,\alpha_n\beta_n}h_{\alpha_1\beta_1}\times
\cdots \times h_{\alpha_n\beta_n}.
\end{equation}
\\
Given the above, we can integrate-out the graviton and the auxiliary field to obtain an effective two-body action
\begin{equation}
e^{i\mathcal S_{\rm eff}[x_l,x_H]}=\int \mathcal{D}h\ \mathcal{D}A\
e^{i\mathcal S_{EH}[h,A]+i \mathcal S_{GF}[h]+i\mathcal
  S_l[x_l,h]+i\mathcal S_H[x_H,h]}.
\end{equation}
The self-force effective action has an expansion in  powers $m/M\ll1$ 
\begin{equation}\label{e:Seffexp}
\mathcal S_{\rm eff}=-\frac{M}{2}\int d\tau_H\ \eta^{\mu\nu}v_{H\mu}v_{H\nu}+M\sum_{n=0}^{\infty}\int d\tau_l\
\left(\frac{m}{M}\right)^{n+1}\mathcal{L}_n[x_l(\tau_l),x_H(\tau_H)],
\end{equation}
where the leading is the worldline action for the heavy body.
We parametrize the trajectory of the light body as
\begin{equation}
    x_l^{\mu}(\tau_l)\equiv x^{\mu}(\tau)=\sum_{n=0}^{\infty}\left(\frac{m}{M}\right)^n\delta
    x^{(n)\,\mu}(\tau),
  \end{equation}
  and the one for the heavy body as
\begin{equation}
x_H^{\mu}(\tau_H)=u_H^{\mu}\tau_H+\sum_{n=1}^{\infty}\left(\frac{m}{M}\right)^n\delta
x_H^{(n)\,\mu}(\tau_H),
\end{equation}
where $\frac{d}{d\tau}\delta x^{(n)\,\mu}(\tau)=\delta
v^{(n)\,\mu}(\tau)$ and $u^{\mu}_H=(-1,0,0,0)$ as we are working in the rest frame of the heavy body.  Notice that the first correction to the trajectory of the heavy body starts at the order $m/M$, so that for the geodesic we can use $x_H^\mu(\tau_H)=u_H^\mu \,\tau_H$. The leading order trajectory of the light body can be derived from the equations-of-motion with respect to the 0SF effective action $\mathcal{L}_0[x(\tau),u_H^{\mu}\tau_H]$.  
The higher-order self-force contributions are obtained by inserting the $m/M$ expansion of  the worldline trajectories  in~\eqref{e:Seffexp}.
Therefore, at a given order in the expansion, $\mathcal{L}_n[x(\tau),x_H(\tau_H)]$ is evaluated on the leading order $(m/M)^0$ worldline trajectories supplemented by 
 $\mathcal{L}_{r<n}$ evaluated after expanding the worldline trajectories to the corresponding order in the self-force expansion.
  Finally, we should note that we haven't specified the kinematics of the system, therefore the formalism is suitable both for the bound and the scattering problem.

\section{Geodesic motion from leading order effective action}\label{sec:geodesic}

 In order to derive the geodesic equation for the light body, one needs
to compute
$\mathcal{L}_0[x^{\mu}(\tau),u^{\mu}_H\tau_H]$, namely the leading order effective action evaluated on the zeroth order trajectory of the heavy body. This corresponds to the leading self-force (0SF), in the expansion in $m/M$ which includes all orders in $G_N$,
because  it generates  the geodesic equation for the light body
in the exact Schwarzschild background of the heavy body, as it will be shown below.
For the present 0SF derivation only the  off-shell graviton emission
from the heavy body is required since there is no worldline vertex for the auxiliary field.

\medskip

It was shown in~\cite{Mougiakakos:2024nku} that it
is possible to derive to all order in $G_N$ the ``\textit{dressed}''
off-shell currents for the graviton by taking the static limit of the heavy body

\begin{equation}
J_{\mu\nu}(\textbf{k})=\begin{tikzpicture}[baseline={([yshift=-.5ex]current bounding box.center)},scale=1]
\newcommand\Square[1]{+(-#1,-#1) rectangle +(#1,#1)}
      \draw[boson] (0,-0.5) -- (0.0,-1.5);
     \draw [fill] (0,-1.5) \Square{3pt} [radius=2pt];
   \end{tikzpicture}=\sum_{n=1}^{\infty}J_{\mu\nu}^{(n)}(\textbf{k}).
\end{equation}
Focusing on four dimensions $D=4$, the expressions are known to all order
\begin{equation}
J^{(n)}_{\mu\nu}(\textbf{k})=\rho(|\textbf{k}|,n) \,\left(\chi_1^{(n)}\delta_{\mu}^0\delta_{\nu}^0+\chi_2^{(n)}\left(\eta_{\mu\nu}-\frac{k_{\mu}k_{\nu}}{\textbf{k}^2}\right)\right),
\end{equation}
where $\textbf{k}=\sum_{i=1}^3 (k_i)^2$ and  in $D=4$ the form factors for the graviton vertices are given in eqs.~(23),~(43) and~(44) of~\cite{Mougiakakos:2024nku}
\begin{eqnarray}\label{e:chires}
  \chi_1^{(1)}&=4, \quad \chi_1^{(2)}={15\over2}, \quad \chi_1^{(n)}&=8 \qquad
                \textrm{for}\qquad n\geq3,\cr
                \chi_2^{(1)}&=0,  \quad \chi_2^{(2)}={1\over2}, \quad \chi_2^{(n)}&=0 \qquad
  \textrm{for}\qquad n\geq3.
\end{eqnarray}
Finally the expansion parameter reads
\begin{equation}\label{e:rhodef}
  \rho(|\textbf{k}|,n)=\frac{(2\sqrt{\pi})^3\Gamma\left(\frac{3-n}{2}\right)}{2^n\Gamma\left(\frac{n}{2}\right)}\,\frac{\left(
    G_NM\right)^n}{|\textbf{k}|^{3-n}}.
\end{equation}

The zeroth order effective action $\mathcal{L}_0[x^{\mu}(\tau),u^{\mu}_H\tau_H]$ is built by summing the exchange
between the light body, represented by a black circle, and the heavy body,
represented by a black square
\begin{equation}\label{e:effac}
    \mathcal{L}_0[x^{\mu}(\tau),u^{\mu}_H\tau_H]=
   \begin{tikzpicture}[baseline={([yshift=-.5ex]current bounding box.center)},scale=1]
    \newcommand\Square[1]{+(-#1,-#1) rectangle +(#1,#1)}
      \draw [fill] (0,0) circle [radius=2pt];
    \end{tikzpicture}
    +
    \begin{tikzpicture}[baseline={([yshift=-.5ex]current bounding box.center)},scale=1]
    \newcommand\Square[1]{+(-#1,-#1) rectangle +(#1,#1)}
      \draw[boson] (0,-1.5) -- (0,0);
      \draw [fill] (0,0) circle [radius=2pt];
        \draw [fill] (0,-1.5) \Square{3pt} [radius=2pt];
    \end{tikzpicture}+
    \begin{tikzpicture}[baseline={([yshift=-.5ex]current bounding box.center)},scale=1]
    \newcommand\Square[1]{+(-#1,-#1) rectangle +(#1,#1)}
      \draw[boson] (0.5,-1.5) -- (0,0);
      \draw[boson] (-0.5,-1.5) -- (0,0);
      \draw [fill] (0,0) circle [radius=2pt];
        \draw [fill] (0.5,-1.5) \Square{3pt} [radius=2pt];
        \draw [fill] (-0.5,-1.5) \Square{3pt} [radius=2pt];
    \end{tikzpicture}+
    \begin{tikzpicture}[baseline={([yshift=-.5ex]current bounding box.center)},scale=1]
    \newcommand\Square[1]{+(-#1,-#1) rectangle +(#1,#1)}
      \draw[boson] (0.5,-1.5) -- (0,0);
      \draw[boson] (-0.5,-1.5) -- (0,0);
      \draw[boson] (0,-1.5) -- (0,0);
      \draw [fill] (0,0) circle [radius=2pt];
        \draw [fill] (0.5,-1.5) \Square{3pt} [radius=2pt];
        \draw [fill] (-0.5,-1.5) \Square{3pt} [radius=2pt];
        \draw [fill] (0,-1.5) \Square{3pt} [radius=2pt];
    \end{tikzpicture}+
    \begin{tikzpicture}[baseline={([yshift=-.5ex]current bounding box.center)},scale=1]
    \newcommand\Square[1]{+(-#1,-#1) rectangle +(#1,#1)}
      \draw[boson] (0.5,-1.5) -- (0,0);
      \draw[boson] (-0.5,-1.5) -- (0,0);
      \draw[boson] (0.15,-1.5) -- (0,0);
      \draw[boson] (-0.15,-1.5) -- (0,0);
      \draw [fill] (0,0) circle [radius=2pt];
        \draw [fill] (0.5,-1.5) \Square{3pt} [radius=2pt];
        \draw [fill] (-0.5,-1.5) \Square{3pt} [radius=2pt];
        \draw [fill] (-0.15,-1.5) \Square{3pt} [radius=2pt];
        \draw [fill] (0.15,-1.5) \Square{3pt} [radius=2pt];
    \end{tikzpicture}+\cdots
\end{equation}
The black squares denote the \emph{exact} Schwarzschild metric generated by
the heavy body, and the wavy lines are for  the emission of gravitons.
Therefore, the evaluation of the
$\mathcal{L}_0[x^{\mu}(\tau),u^{\mu}_H\tau_H]$ is obtained by a double
resummation:\\

(1)
A first summation of the all loop interactions, denoted by $L$,  between the light body
worldline with the Schwarzschild background
\begin{equation}\label{e:effacexp}
 \mathcal{L}_0[x^{\mu}(\tau),u^{\mu}_H\tau_H] =\frac{1}{2}v_{\mu}(\tau)v_{\nu}(\tau)\left(-\eta^{\mu\nu}+\sum_{L=1}^{\infty}\mathcal{I}^{(L)}_{\alpha_1\beta_1,\dots,\alpha_L\beta_L}\mathcal{T}^{\mu\nu\alpha_1\beta_1,\dots,\alpha_L\beta_L}_{(L)}\right),
\end{equation}
where
$\mathcal{T}^{\mu\nu\alpha_1\beta_1,\dots,\alpha_L\beta_L}_{(L)}$ are
defined in~\eqref{e:proj}. 

(2) At each loop orders,  the
$\mathcal{I}^{(L)}_{\alpha_1\beta_1,\dots,\alpha_L\beta_L}$ tensors are given by 
 infinite sums building the Schwarzschild metric generated by the heavy body~\cite{Mougiakakos:2024nku}
\begin{multline}\label{e:Ilexpr}
\mathcal{I}^{(L)}_{\alpha_1\beta_1,\dots,\alpha_L\beta_L}=
\int_{\mathbb R^3} {d^3\textbf{k} \over (2\pi)^3}  e^{i\textbf{k}\cdot \textbf{x}(\tau)} \int_{\mathbb R^{3L}} \prod_{i=1}^L {d^3 \textbf{q}_i\over (2\pi)^3}\delta^{(3)}\left(\sum_{i=1}^L\textbf{q}_i-\textbf{k}\right)\cr
\times\sum_{n_1=1}^{\infty}\cdots\sum_{n_L=1}^\infty\rho(|\textbf{q}_i|,n_i) \,\left(\chi_1^{(n_i)}\delta_{\alpha_i}^0\delta_{\beta_i}^0+\chi_2^{(n_i)}\left(\eta_{\alpha_i\beta_i}-\frac{q_{i,\alpha_i}q_{i,\beta_i}}{\textbf{q}_i^2}\right)\right),
\end{multline}
where $\rho(|\textbf{q}_i|,n_i) $ is defined in~\eqref{e:rhodef} and
the metric form factors are given in~\eqref{e:chires}.\\

\medskip
Using the integral representation for  the delta-function of the
momentum conservation
\begin{equation}
  \delta^{(3)}\left(\sum_{i=1}^L \textbf{q}_i-\textbf{k}\right)= \int_{\mathbb R^3}  e^{i
    \textbf{u}\cdot \left(  \sum_{i=1}^L
      \textbf{q}_i-\textbf{k}\right)} {d^3\textbf{u}\over(2\pi)^3}
\end{equation}
we obtain that
\begin{multline}\label{e:Ildecoupled}
\mathcal{I}^{(L)}_{\alpha_1\beta_1,\dots,\alpha_L\beta_L}=\int_{\mathbb R^{3L}}
\prod_{i=1}^L {d^3 \textbf{q}_i\over (2\pi)^3} e^{i
  \textbf{x}(\tau)\cdot \textbf{q}_i}\cr\times
\sum_{n_1,\dots,n_L=1}^{\infty}\rho(|\textbf{q}_i|,n_i) \,\left(\chi_1^{(n_i)}\delta_{\alpha_i}^0\delta_{\beta_i}^0+\chi_2^{(n_i)}\left(\eta_{\alpha_i\beta_i}-\frac{q_{i,\alpha_i}q_{i,\beta_i}}{\textbf{q}_i^2}\right)\right).
\end{multline}
The Fourier transform to direct space has decoupled the multiloop integrals, and
we just have to evaluate the following two type of integrals
\begin{equation}
 \int_{\mathbb R^3} {d^3\textbf{q}\over(2\pi)^3}
 {e^{i\textbf{x}\cdot\textbf{q}}\over
   |\textbf{q}|^{3-n}}=-{\Gamma(n-1)\over |\textbf{x}|^n} {\cos\left(n\pi\over2\right)\over2\pi^2},
\end{equation}
and for $i,j=1,2,3$ 
\begin{equation}
 \int {d^3\textbf{q}\over(2\pi)^3}
 {e^{i\textbf{x}\cdot\textbf{q}}\over |\textbf{q}|^{3-n}}
\left (\delta_{ij}-{q_i q_j\over
    \textbf{q}_i^2}\right)=
-{\Gamma(n-1)\over |\textbf{x}|^n }{\cos\left(n\pi\over2\right)\over2\pi^2} \,\left(\delta_{ij}{n-2\over n-3}-{n\over n-3} n_i
  n_j\right) 
\end{equation}
where $n_i=x_i/|\textbf{x}|$ is the spatial unit vector.
Collecting everything and summing over $n_1,\dots,n_L$, we obtain the simple form 

\begin{equation}\label{result}
\mathcal{I}^{(L)}_{\alpha_1\beta_1,\dots,\alpha_L\beta_L}=\rho^{2L}\prod_{i=1}^L\left(\delta^0_{\alpha_i}\delta^0_{\beta_i}
  \,\left(\frac{4(1+\rho)}{\rho(1-\rho)}-1\right)+n_{\alpha_i}n_{\beta_i}\right)
\end{equation}
with
\begin{equation}
  \rho\equiv\rho(|\textbf{x}|(\tau))=\frac{G_N M}{|\textbf{x}|(\tau)}, 
\end{equation}
since we Fourier transformed with respect to the position $\textbf{x}(\tau)$ of the light body.
One can easily observe the recursive relation
\begin{equation}
\mathcal{I}^{(L)}_{\alpha_1\beta_1,\dots,\alpha_L\beta_L}=\mathcal{I}^{(L-1)}_{\alpha_1\beta_1,\dots,\alpha_{L-1}\beta_{L-1}}\rho^2\left(\delta^0_{\alpha_L}\delta^0_{\beta_L}\left(\frac{4(1+\rho)}{\rho(1-\rho)}-1\right)+n_{\alpha_L}n_{\beta_L}\right).
\end{equation}
stating that the $L$-loop contribution to~\eqref{e:effac} is obtained
by adjoining an extra graviton line.
Using the properties of the projector in~\eqref{e:proj} one can find
another useful recursion relation
\begin{multline}
\mathcal{I}^{(L)}_{\alpha_1\beta_1,\dots,\alpha_L\beta_L}\mathcal{T}_{(L)}^{\mu\nu\alpha_1\beta_1,\dots,\alpha_L\beta_L}=-\eta^{\mu\nu}\mathcal{I}^{(L)}_{\alpha_1\beta_1,\dots,\alpha_L\beta_L}\mathcal{P}_{(L)}^{\alpha_1\beta_1,\dots,\alpha_L\beta_L}\cr
+\mathcal{I}^{(L-1)}_{\alpha_1\beta_1,\dots,\alpha_{L-1}\beta_{L-1}}\mathcal{P}_{(L-1)}^{\alpha_1\beta_1,\dots,\alpha_{L-1}\beta_{L-1}}\rho^2\left(\delta^{\mu}_0\delta^{\nu}_0  \left(\frac{4(1+\rho)}{\rho(1-\rho)}-1\right)+n^{\mu}n^{\nu}\right),
\end{multline}
where $\mathcal{P}_{(n)}^{\alpha_1\beta_1,...,\alpha_n\beta_n}$ is
defined from the expansion of the determinant of the gothic metric
in~\eqref{e:proj2}. Taking a closer look at this expression in $D=4$
we have
\begin{equation}\label{e:DetId}
  \left(-\text{det}[\eta^{\mu\nu}-\sqrt{32\pi
    G_N}h^{\mu\nu}]\right)^{-1/2}=1+\sum_{n=1}^{\infty} (32\pi
    G_N)^{n\over2} \mathcal{P}_{(n)}^{\alpha_1\beta_1,\dots,\alpha_n\beta_n}h_{\alpha_1\beta_1}\cdots
h_{\alpha_n\beta_n}.
\end{equation}
Therefore, using~\eqref{result}, we have that
\begin{equation}\label{e:Isum}
\sum_{L=1}^{\infty}\mathcal{P}_{(L)}^{\alpha_1\beta_1,\dots,\alpha_L\beta_L}\mathcal{I}^{(L)}_{\alpha_1\beta_1,\dots,\alpha_L\beta_L}=\left(-\text{det}\left[\eta^{\mu\nu}-\rho^2\left(\delta_0^{\mu}\delta_0^{\nu}\left(\frac{4(1+\rho)}{\rho(1-\rho)}-1\right) +n^{\mu}n^{\nu}\right)\right]\right)^{-\frac{1}{2}}-1
\end{equation}
which evaluates to 
\begin{equation}
\sum_{L=1}^{\infty}\mathcal{P}_{(L)}^{\alpha_1\beta_1,\dots,\alpha_L\beta_L}\mathcal{I}^{(L)}_{\alpha_1\beta_1,\dots,\alpha_L\beta_L}=\frac{1}{(1+\rho)^2}-1.
\end{equation}
Plugging  back this result  in~\eqref{e:Isum} leads to
\begin{equation}
\sum_{L=1}^{\infty}\mathcal{I}^{(L)}_{\alpha_1\beta_1,\dots,\alpha_L\beta_L}\mathcal{T}_{(L)}^{\mu\nu\alpha_1\beta_1,\dots,\alpha_L\beta_L}=\frac{\rho}{(1+\rho)^2}\left(\delta^{\mu}_0\delta^{\nu}_0\left(\frac{4(1+\rho)}{(1-\rho)}-\rho\right)+\left(\rho+2\right)\eta^{\mu\nu}+ \rho\ n^{\mu}n^{\nu}\right),
\end{equation}
and using~\eqref{e:effacexp}, we get 
\begin{equation}
\mathcal{L}_0[x^{\mu}(\tau),u^{\mu}_H\tau_H]=\frac{v_{\mu}(\tau)v_{\nu}(\tau)}{2}\frac{1}{(1+\rho)^2}\left(\delta^{\mu}_0\delta^{\nu}_0\left(\frac{4\rho(1+\rho)}{(1-\rho)}-\rho^2\right)-\eta^{\mu\nu}+\rho^2\ n^{\mu}n^{\nu}\right).  
\end{equation}

We are ready to compute
the corresponding equations-of-motion of the light body at 0SF
order. Before doing so, we can simplify the notation by observing that
in the harmonic gauge, used in this work, the inverse metric takes the form\footnote{This is obtained  by computing the inverse metric from eq.~(5.3) of~\cite{Mougiakakos:2020laz} and using $f(\textbf{x})=1+\rho$ and $d=3$
\begin{equation}\label{e:SchwaF}
ds^2=-h_0(\textbf{x}) dt^2+  h_1(\textbf{x}) d\vec x^2+h_2(\textbf{x}) {(\vec           x\cdot
  d\vec x)^2\over \textbf{x}^2},
\end{equation} 
with 
\begin{equation}
\label{e:hfinitedef} h_0(r):=1- {2\rho\over f(\textbf{x})}, \,
h_1(r):=  f(\textbf{x})^2, \,
 h_2(r):=-f(\textbf{x})^2+f(\textbf{x}) {(f(\textbf{x})+\textbf{x} {df(\textbf{x})\over d\textbf{x}})^2\over
                  f(\textbf{x})-2\rho }\,.
\end{equation} 
These equations are converted to the mostly plus metric used in this paper, and we correct a typographical error in the expression for $h_2(\textbf{x})$.}
\begin{equation}
 g^{\mu\nu}(\rho)=-\frac{1}{(1+\rho)^2}\left(\delta^{\mu}_0\delta^{\nu}_0\left(\frac{4\rho(1+\rho)}{(1-\rho)}-\rho^2\right)-\eta^{\mu\nu}+\rho^2\ n^{\mu}n^{\nu}\right)
\end{equation}
so that we can actually write the derived leading order effective action as
\begin{equation}
\mathcal{L}_0[x^{\mu}(\tau),u^{\mu}_H\tau_H]=-\frac{1}{2}v_{\mu}(\tau)v_{\nu}(\tau)g^{\mu\nu}(|\textbf{x}|(\tau)),
\end{equation}
where $g^{\mu\nu}$ is the inverse metric in harmonic coordinates. In this notation, it is easy
to see that  the Lagrangian for the light body reads 
\begin{equation}
\mathcal{L}_l    = m\ \mathcal{L}_0+\mathcal{O}(m/M)
\end{equation}
evaluated at the rest frame of the heavy body and the equations-of-motion for the light body derived in the self-force EFT formalism  are  exactly equivalent to the geodesic equation for a massive probe in a Schwarzschild spacetime.

\section{Discussion}\label{sec:conclusion}

We have presented a scattering amplitude-based derivation of the exact geodesic for a light point-like body in the background of a heavy point-like body. The geodesic is obtained by summing two infinite families of contributions: 
(1) The gravitational contributions generated by the heavy body are given by the sum over $n_1,\dots,n_L$ in equation~\eqref{e:Ilexpr}---this sum generates the Schwarzschild black hole metric derived in reference~\cite{Mougiakakos:2024nku}---; (2) The sum  over all  the graviton exchanges between the heavy body and light body, as described in~\eqref{e:effacexp}.

This double infinite summation captures all the probe contributions of the post-Minkowskian expansion around a flat spacetime considered in~\cite{Damour:2017zjx,Brandhuber:2021eyq,Bjerrum-Bohr:2021wwt}. 
In order to perform the summation of the probe contributions, it is necessary to have knowledge of the tensors $\mathcal{P}_{(n)}^{\alpha_1\beta_1,\dots,\alpha_n\beta_n}$ from equation~\eqref{e:proj2} to all orders in $G_N$. This is a technically challenging task. Fortunately, the identity in Equation~\eqref{e:DetId} allows  performing the sum over all loop contributions without the need to explicitly determine these tensors. It would be of interest to identify such simplifications in the post-Minkowskian-based approach.

Another simplification occurred as a consequence of the Fourier transform to position space, which decoupled the loop integrals in equation~\eqref{e:Ildecoupled}.  In fact, at each loop order, the tensor $\mathcal{I}^{(L)}_{\alpha_1\beta_1,\dots,\alpha_L\beta_L}$ reduces to a product of simple Fourier integrals. The identity in Equation~\eqref{e:DetId} shows that one just need to replace the $\mathcal{I}^{(L)}_{\alpha_1\beta_1,\dots,\alpha_L\beta_L}$ with  $\prod_{i=1}^L h_{\alpha_i\beta_i}$ from the  expansion of the metric in position space computed in~\cite{Mougiakakos:2024nku}. Consequently, in practice, the intricate loop integrals that must be evaluated are ultimately straightforward.  

The simplicity of deriving the geodesic from the scattering amplitude
in the cubic formulation of gravity is another illustration of the
fact that the complexity of the computation depends heavily on the
chosen formulation. The fact that the all order summation is possible in this cubic formulation, contrary to previous attempts~\cite{Mougiakakos:2020laz}, is a strong indicator that there exists a more suitable field basis for the metric degrees-of-freedom. This is consistent with the EFT spirit, in the sense that the metric degrees-of-freedom, formally integrated-out, serve only as an intermediate step in the computation of the two-body effective action which is invariant under these choices.
Consequently, the cubic formulation of gravity and the worldline formalism adapted for a self-force expansion is a promising avenue towards the direction of a complete self-force EFT computation to higher orders.

 \section{Acknowledgements}
We would like to thank R. Aoude and D. Kosmopoulos for useful comments and discussions. S.M. acknowledges financial support by ANR PRoGRAM project, grant ANR-21-CE31-0003-001.

\end{document}